\documentclass[twocolumn,floatfix,prl,showpacs]{revtex4-1}
\usepackage{graphicx}
\usepackage{amssymb}
\usepackage{amsmath}
\usepackage{color}
\usepackage{psfrag}
\usepackage{epsfig}
\usepackage{dsfont}
\usepackage{bbm}
\usepackage{bm}

\newcommand{\ket}[1]{| #1 \rangle}
\newcommand{\bra}[1]{\langle #1 |}
\newcommand{\rb}[1]{\left( #1 \right)}

\newcommand{\ew}[1]{\langle #1 \rangle}
\newcommand{\beq}{\begin{eqnarray}}
\newcommand{\eeq}{\end{eqnarray}}

\newcommand{\op}[2]{| #1 \rangle \langle #2 |}

\newcommand{\eq}[1]{Eq.~(\ref{#1})}
\newcommand{\fig}[1]{Fig.~\ref{#1}}

\newcommand{\trace}[1]{\mathrm{Tr}\left\{#1\right\}}

\begin{document}
\title{Leggett-Garg inequalities for the statistics of electron transport}
\author{Clive Emary}
\affiliation{
  Institut f\"ur Theoretische Physik,
  Hardenbergstr. 36,
  TU Berlin,
  D-10623 Berlin,
  Germany
}

\date{\today}
\begin{abstract}
  We derive a set of Leggett-Garg inequalities (temporal Bell's inequalities) for the moment generating function of charge transferred through a conductor.  Violation of these inequalities demonstrates the absence of a macroscopic-real description of the transport process.
  We show how these inequalities can be violated by quantum-mechanical systems and consider transport through normal and superconducting single-electron transistors as examples.
\end{abstract}
\maketitle


Full counting statistics (FCS) seeks to understand electronic transport by counting the number of charges transferred through a conductor in a certain time interval $t_b\ge t\ge t_a$\cite{Levitov1996,*Levitov2002}.  
Considered as a {\em classical} stochastic process, the information about transferred charge can be encapsulated by the moment generating function (MGF)
\beq
  \mathcal{G}_\mathrm{cl.}(\chi;t_b,t_a) = \ew{e^{i\chi[ n(t_b) - n(t_a) ]}}
  \label{MGFcl}
  ,
\eeq
where $n(t)$ is the collector charge at time $t$ and $\chi$ is the counting field.
Recent quantum dot (QD) experiments, \cite{Gustavsson2006,Fujisawa2006}, have borne out many of the predictions of FCS 
in a regime where charge transfer is essentially classical and 
the above definition applicable.
More generally though, the collector charge is a quantum degree of freedom since electrons can form superpositions between states within the reservoir and those without.

The aim of this paper is is to investigate the dividing line between classical and quantum transport by deriving a set of inequalities for the MGF.  These inequalities are obeyed by all classical systems but, as we go on to show, can be violated by quantum-mechanical ones.

In contrast to the more familiar spatial Bell's inequalities that probe entanglement between particles \cite{Bell1964,Clauser1969} and have been extensively discussed in the context of quantum transport, e.g. \cite{Chtchelkatchev2002,Samuelsson2003,Beenakker2003,Emary2009, Bednorz2011}, the inequalities we derive here are of the class introduced by Leggett and Garg --- single-system temporal Bell's inequalities \cite{Leggett1985,*Leggett2002}, which have been the subject of several recent (non-transport) experiments, e.g. \cite{Palacios-Laloy2010,Xu2011}.   
Here we write down inequalities for the MGF of FCS which are predicated on the {\em macroscopic reality} \cite{Leggett1985,*Leggett2002} of the collector charge --- classically, the charge has a definite (if unknown) value at all times, which can, in principle, be measured non-invasively.    
This type of inequality was considered for transport systems in Ref.~\cite{Lambert2010}, but the focus there was on the the internal degrees-of-freedom of system, and not the FCS as measured in the contacts, as discussed here.
We may also contrast our work with that of Vogel at al. \cite{Vogel2000,Richter2002} who derived inequalities for the generating function of position for a harmonic oscillator.

We show that our inequalities can be violated by the quantum transport.  As examples, we consider two single electron transistors (SETs): first, a normal SET for which the violation is limited; and secondly, a superconducting-SET (SSET), for which the violations are much more pronounced. This latter result clearly demonstrates that the double-Josephson quasi-particle resonance (DJQP) phenomenon of the SSET \cite{Hadley1998,Clerk2002, *Clerk2003,Thalakulam2004,Xue2009,Kirton2012} has no macroscopic-real interpretation.

\section{Inequalities}
The inequalities that we derive here concern the quantity
\beq
  L(\chi,\left\{t_i\right\})
  \equiv
  \mathcal{G}(\chi;t_1,t_0) 
  + 
  \mathcal{G}(\chi;t_2,t_1)
  -
  \mathcal{G}(\chi;t_2,t_0)
  \label{Ldefn}
  ,
\eeq 
which involves the MGF over three different time intervals.  To derive classical bounds for this quantity we begin by writing the classical MGF of \eq{MGFcl} as
\beq
  \mathcal{G}_\mathrm{cl.}(\chi;t_b,t_a) =
  \sum_{n_b, n_a} 
  P\rb{ n_b, n_a}
  e^{i\chi[ n_b - n_a ]}
  \label{MGFclP}
\eeq
where $P\rb{ n_b, n_a}$ is the probability of having $n_a = n(t_a)$ collector charges electrons at time $t_a$ and $n_b=n(t_b)$ at time $t_b$.
Under the Leggett-Garg assumptions of macroscpic realism and non-invasive measurability \cite{Leggett1985}, the three probabilities required to construct the classical expression 
$L_\mathrm{cl.}(\chi,\left\{t_i\right\})$ of \eq{Ldefn} can be obtained as marginals of the joint probability $P\rb{ n_3,n_2, n_1}$, e.g. 
$P\rb{ n_3, n_1} = \sum_{n_2}P\rb{ n_3,n_2, n_1}$.
This allows us to write \eq{Ldefn} classically as
\beq
  L_\mathrm{cl.}
  &=&
  \sum_{\left\{n_i\right\}} 
  P\rb{ n_3,n_2, n_1}
  \left\{
  e^{i\theta_{10}}
  +e^{i\theta_{21}}
  -
  e^{i\theta_{20}}
  \right\}
,
\eeq 
where we have introduced the shorthand $\theta_{ba} = \chi[ n_b - n_a ]$.  Taking the real part, we have
\beq
  \mathrm{Re}\left\{
    L_\mathrm{cl.}
    \right\}
  &=&
  \sum_{\left\{n_i\right\}} 
  P\rb{ n_3,n_2, n_1}
  \nonumber\\
  &&
  ~~~~~
  \times
  \left\{
  \cos (\theta_{10}) + \cos (\theta_{21}) 
    - \cos (\theta_{20})
  \right\}
  \nonumber
  ,
\eeq 
An upper bound for this quantity is obtained by finding the maximum value of the quantity in brackets 
as a function of electron numbers $\left\{n_i\right\}$ and choosing the probability distribution $P\rb{ n_3,n_2, n_1}$ such that all weight resides with this maximum value. We have then
 \beq
  \mathrm{max}
  \left[
    \mathrm{Re}\left\{L_\mathrm{cl.}\right\}
  \right]
  &=&
  \mathrm{max}
  \left[
    \cos (\theta_{10}) + \cos (\theta_{21}) 
  \right.
  \nonumber\\
  &&
  ~~~~~~~~~~~~~~~~~
  \left.
      - \cos (\theta_{21}+\theta_{10})
  \right]
  \nonumber
  ,
\eeq 
where we have used $\theta_{20} = \theta_{21} + \theta_{10}$.
The maximum on the righthand-side can then be found solving $ \sin(\theta_{10}) = \sin(\theta_{21}) = \sin (\theta_{21} + \theta_{10})$ subject to the constraints that each $\theta_{ij}$ is equal to $\chi$ times an integer.  These constraints arise from the quantisation of the collector charge in integer units. We thus obtain an upper bound for $\mathrm{Re}\left\{L_\mathrm{cl.}\right\}$ that we will denote  $C_\mathrm{R}(\chi)$.  This bound is $\chi$-dependent and selected values are listed in Table~\ref{Ctab}.  A lower bound $B_\mathrm{R}(\chi)$ can similarly be established but in the examples we study here, this lower bound is not violated and we will not consider it further.
Using a similar argument, the imaginary part of $L_\mathrm{cl.}$ can also be shown to be bounded, from above by $C_\mathrm{I}(\chi)$ and below by $-C_\mathrm{I}(\chi)$(Table \ref{Ctab}).  
The central formal result of this paper is therefore that any classical MGF must obey the inequalities:
\begin{subequations}
\label{ineq}
\beq
  \mathrm{Re}\!
  \left\{\frac{}{}\!\!
  \mathcal{G}(\chi;t_1,t_0) 
     + 
     \mathcal{G}(\chi;t_2,t_1)
     -
     \mathcal{G}(\chi;t_2,t_0)
     \!
  \right\}
  \le C_\mathrm{R}(\chi)
  ;
  ~~~~
  \label{ineqR}
 \\
 \left|
     \!
  \mathrm{Im}\!\!
  \left\{\frac{}{}\!\!
  \mathcal{G}(\chi;t_1,t_0) 
     + 
     \mathcal{G}(\chi;t_2,t_1)
     -
     \mathcal{G}(\chi;t_2,t_0)
     \!
  \right\}
     \!
  \right|
  \le C_\mathrm{I}(\chi)
  ,
  ~~~~
  \label{ineqI}
\eeq
\end{subequations}
for all $\chi$ and times $\left\{t_i\right\}$.

For simplicity, in the following we will set $t_0=0$, $t_1=\tau$ and $t_2=2\tau$and define
\beq
  R(\chi,\tau) 
  \equiv\mathrm{Re}[
  \mathcal{G}(\chi;\tau,0) 
     + 
     \mathcal{G}(\chi;2\tau,\tau)
     -
     \mathcal{G}(\chi;2\tau,0)
     ]
  ,
\eeq
with $I(\chi,\tau)$ as the imaginary part defined analogously.
In the stationary limit, the MGFs are translationally invariant and we have, e.g.,
$
  R(\chi,\tau)
  \equiv
  \mathrm{Re}
  [
    2\mathcal{G}(\chi;\tau,0)
     -
     \mathcal{G}(\chi;2\tau,0)
  ]
$.

\begin{table}[t]
  \begin{tabular}{|c|ccccc|}
    \hline
     $\chi$ & 
       $\frac{\pi}{4}$ & $\frac{\pi}{3}$ & 
       $\frac{\pi}{2}$ & $\frac{2\pi}{3}$ 
       & $\pi$ 
     \\
    \hline
     $C_\mathrm{R}(\chi)$ 
       & $\sqrt{2}$     &  $\frac{3}{2}$ 
       & $1$  &  $1$ 
       & $1$ \\
    $B_\mathrm{R}(\chi)$ 
       & $-3$     &  $-3$ 
       & $-3$  &  $-2$ 
       & $-3$ \\
     $C_\mathrm{I}(\chi)$ 
       & $1+\sqrt{2}$  & $\frac{3\sqrt{3}}{2}$ 
       & $2$ & $\frac{3\sqrt{3}}{2}$ 
       & $0$ \\
    \hline
  \end{tabular}
  \caption{
  Bounds of \eq{ineq} for selected values of $\chi$.  $C_\mathrm{R}(\chi)$ and $B_\mathrm{R}(\chi)$ are upper and lower bounds for the real part inequality, \eq{ineqR}; $C_\mathrm{I}(\chi)$ bounds the madnitude of the imaginary part, \eq{ineqI}.
  \label{Ctab}
  }
\end{table}

\section{Quantum MGF}
Quantum-mechanically, there is no unique generalisation of \eq{MGFcl} since the MGF is a two-time quantity, and a time-ordering must be specified.  The canonical MGF given by Levitov and coworkers involves Keldysh-ordering \cite{Levitov1996, *Levitov2002}:
\beq
  \mathcal{G}_\mathrm{L}(\chi;t_b,t_a) 
  &=& 
  \ew{
    e^{-i\frac{\chi}{2} \hat{n}(t_a)}
    e^{i\chi\hat{n}(t_b)}
    e^{-i\frac{\chi}{2} \hat{n}(t_a)} 
  }
  \label{MGFlevitov}
  .
\eeq 
The set-up proposed by Levitov et al. to measure this MGF was a spin processing under the influence of the magnetic field generated by the collector current.  This set-up is suited for out purposes because, not only does it yield the complete MGF directly, but this measurement is classically non-invasive.
In a more realistic experimental set-up, it might be the case that only a finite number of cumulants be known, e.g. \cite{Flindt2009}.  In this case, care must be taken in reconstructing the full MGF since the generic behaviour of the cumulants is factorial growth with their order \cite{Flindt2008}.  Care must also be taken that the measurement be performed non-invasively.

In calculating the MGF here, we use the equivalent expression
$
  \mathcal{G}_\mathrm{L}(\chi;t) = \trace{\varrho(\chi;t)}
$ in terms of the $\chi$-resolved density matrix, the dynamics of which are determined by the modified von Neumann equation $\dot \varrho(\chi;t) = - i \left[H(\frac{1}{2}\chi)\varrho(\chi;t) -\varrho(\chi;t) H(-\frac{1}{2}\chi)\right]$ with gauge-transformed Hamiltonian
$H(\chi) =  e^{i\chi\hat{n}} H e^{-i\chi\hat{n}}$.
%

\section{Charge qubit}
\begin{figure}[t]
  \includegraphics[width=\columnwidth,clip=true]{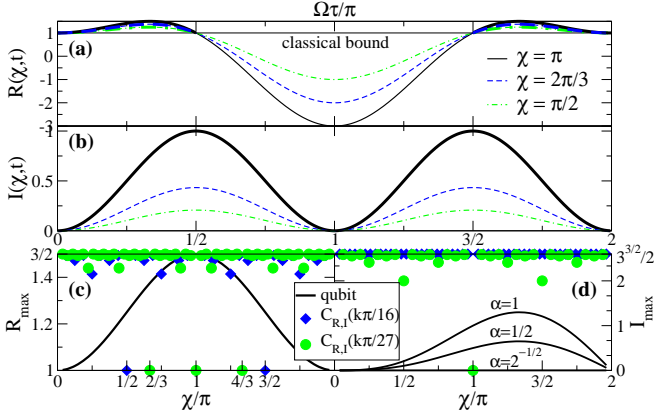}
  \caption{
    (Color online) Violations of \eq{ineq} for an isolated qubit.
    {\bf(a)} and  {\bf(b)} shows the quantities $R(\chi,t)$ and $I(\chi,t)$
    as functions of time for three different values of $\chi/\pi = 1,\frac{2}{3},\frac{1}{2}$.
    Since $C_\mathrm{R}(\chi)=1$ for these values of $\chi$, the qubit evolution violates \eq{ineqR}.  For the imaginary part,  the classical bound is only violated for  $\chi = \pi$ for which $C_\mathrm{I}(\pi)=0$.
    {\bf(c)}    
    The solid line shows the maximum $R_\mathrm{max}(\chi)$ as a function of $\chi$ for the qubit.  The points show the upper classical bound $C_\mathrm{R}(\chi)$ for two sets of $\chi$-values: $\chi = k\pi/2^4$ and $\chi = k\pi/3^3$, with k a non-negative integer. Only for selected values of $\chi$ (notably those used in  {\bf(a)}) is the bound low enough to allow strong violation of the inequality. 
    {\bf(d)}   
    Maximum value $I_\mathrm{max}(\chi)$ and bound $C_\mathrm{I}(\chi)$ as a function of $\chi$. The different qubit lines correspond to different initial conditions.  Only for the special case of $\chi=\pi$ are violations of the classical inequality observed.
    \label{qubitfig}
   }
\end{figure}

We first consider an isolated charge qubit with states $\ket{L}$ and $\ket{R}$,  the latter of which we associate as our ``collector'', such that $\hat{n}=\op{R}{R}$.  We assume a Hamiltonian $H_\mathrm{qb} = \frac{1}{2}\Omega (\op{L}{R}+\op{R}{L})$ with splitting $\Omega$ ($\hbar=1$, here and throughout).
Starting the qubit in an arbitrary state, we find
\beq
  R(\chi,\tau)
     &=& 
     \cos^2 (\textstyle{\frac{1}{2}}\chi) 
     +
     \sin^2 (\textstyle{\frac{1}{2}}\chi) 
     \left\{
     2\cos \Omega\tau
     - \cos 2 \Omega \tau
     \right\}
\eeq
This function is plotted in \fig{qubitfig}a for several different values of $\chi$. Its maximum occurs at a time $t^R_\mathrm{max} = \pi/(3 \Omega)$ and has the value $R_\mathrm{max}(\chi) = 1 + \frac{1}{2}\sin^2(\frac{1}{2}\chi)\le 3/2$. 
Whether or not this constitutes a violation of  \eq{ineq} depends on the value of $C_\mathrm{R}(\chi)$ (see \fig{qubitfig}c).  Maximum violation occurs for $\chi=\pi$ where $R_\mathrm{max}(\pi)= 3/2$ and $C_\mathrm{R}(\pi)=1$.
In this special case, the MGF becomes the two-time correlation function for the  charge-parity operator of the lead, $(-1)^{\hat{n}}$ and \eq{ineqR} reduces to the Legget-Garg inequality for this operator \cite{Kofler2007}.
We note that, although the behaviour of  $R_\mathrm{max}$ is continuous as a function of $\chi$, the bounds, and hence the question of violation, is discontinuous.  

Figure \ref{qubitfig}b shows the imaginary-part $I(\chi,t)$, which does depend on initial conditions. With initial pure state $\alpha\ket{L}+\sqrt{1-\alpha^2}\ket{R}$ with $\alpha$ real, the maximum value reads
$I_\mathrm{max}(\chi)=\ew{\sigma_z}_0(\sin(\frac{1}{2}\chi) - \frac{1}{2}\sin(\chi))$ where $\ew{\sigma_z}_0$ is the expectation value of the Pauli operator $\sigma_z$ in the initial state.

\section{Single electron transistor}
Our first transport example will be the SET with normal leads \cite{Kastner1992}.  We assume strong Coulomb blockade such that only a single spinless level plays a role in transport. The gauge-transformed Hamiltonian of our system is $H(\chi) = H_\mathrm{S} +  H_\mathrm{res} + V(\chi)$, where $ H_\mathrm{S} = \epsilon d^\dag d$ describes the dot level at energy $\epsilon$;  $H_\mathrm{res}= \sum_{k,\alpha} \omega_{k\alpha} c^\dag_{k\alpha} c_{k\alpha}$, two non-interacting leads ($\alpha=L,R$) with states of energy $\omega_{k\alpha}$; and where $V(\chi)= \sum_{k\alpha} t_{k\alpha}e^{i\chi\delta_{\alpha,R}} c^\dag_{k\alpha} d + \mathrm{H.c.}$ describes single-electron tunneling with amplitudes $t_{k\alpha}$.   The leads are taken at temperature $T$ with a  symmetric bias $V$ across the dot such that the chemical potentials are  $\mu_L=-\mu_R=\epsilon \pm eV/2$.   We also assume that the reservoir bands have a Lorentzian cut-off with bandwidth parameter $X_C$. 

We calculate the MGF of \eq{MGFlevitov} by tracing out the leads from the equation-of-motion for $\rho(\chi)$ using the technique described in Ref.~\cite{Marcos2010a,*Emary2011}, and we work to lowest order in the rates  $\Gamma_\alpha(\omega) \equiv 2 \pi \sum_{k} |t_{k\alpha}|^2 \delta(\omega_{k\alpha} - \omega)$, assumed constant.  The resulting {\em nonMarkovian} quantum master equation captures the essential features of system-bath coherence in the limit $\Gamma_\alpha/kT \ll 1$.

\begin{figure}[t]
  \includegraphics[width=\columnwidth,clip=true]{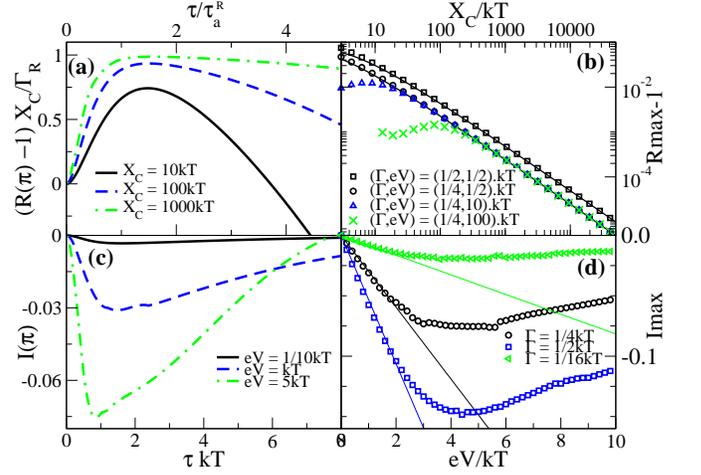}
  \caption{
    (Color online) Violation of inequalities \eq{ineq} with $\chi=\pi$ for the single electron transistor in the sequential tunneling regime ($\Gamma_L=\Gamma_R=\Gamma = \frac{1}{4}kT$, unless stated).
    {\bf (a)} Real part $R(\pi,\tau)-1$ as a function of time $\tau$. Values greater than zero indicate a violation of \eq{ineqR}.  A single maximum is observed at a time $\sim t_a^\mathrm{R}$ (see text) with height inversely proportional to the bandwidth.
    {\bf (b)} Maximum violation of real part inequality as a function of bandwidth. Solid lines correspond to the analytic expression given in the text.
    {\bf (c)} Imaginary part $I(\pi,\tau)$ as function of time $\tau$. 
    Violations occur at a time $\tau \sim 0.7 (\Gamma kT)^{-1/2}$.
    {\bf (d)} Maximum violation of imaginary part inequality as a function of applied bias.  The violation is zero in equilibrium, increases linearly and returns to zero for large bias.
    In calculating $I(\pi,t)$, a large bandwidth $X_C = 10^5kT$ was used.
    \label{SETfig}
   }
\end{figure}

Figure \ref{SETfig} shows that 
despite the weak coupling between system and reservoir, both quantities $R$ and $I$ violate their respective inequalities, albeit in different manners.
The $\chi$-dependence of the results here is similar to that for the qubit and so we concentrate on the case $\chi=\pi$ for which maximum violations occur.
The real-part $R(\pi,\tau)$ shows a single maximum as a function of time (\fig{SETfig}a).  Provided that the bias is less than the bandwidth, $eV \ll X_C$, $R(\pi,\tau)$ is insensitive to the bias.  The maximum occurs at a time
$
  t_\mathrm{max}^R \sim 
  t^R_a = 
  \frac{1}{2X_C} \log \frac{X_C}{\Gamma}
$ and the value of the function at this point is 
$R_\mathrm{max}(\pi) \sim 1 + a (1-\sqrt{a})^2$; $a = \Gamma / X_C$ (\fig{SETfig}b). 
Violation of \eq{ineq} for the SET in sequential regime is therefore an equilibrium effect and relies on finite bandwidth: as $X_C$ increases, the degree of violation (as well as the time of violation) tend to zero:  $R_\mathrm{max}(\pi) -1 \sim  \Gamma / X_C \to 0 $ in wideband limit.

Figure \ref{SETfig}c and d show the imaginary-part $I(\pi,t)$ for this model. In contrast to the real part, this quantity shows violations of the classical inequality in the large bandwidth limit, $X_C \to \infty$.  As a function of time, $I(\pi,t)$ shows a single maximum located 
at
$
  t
  \sim
  \Gamma^{-1} \log 
  \left[
    1+ c + \sqrt{c(1+c)}
  \right]
$ with $c =14 \zeta(3)/\pi^3 \frac{\Gamma}{kT}\approx 0.54 \frac{\Gamma}{kT}$.
The maximum degree of violation, $I_\mathrm{max}(\chi)$, is shown in \fig{SETfig}d.  To lowest order in the bias, we find $I_\mathrm{max}(\pi) \sim - \frac{1}{4} eV \frac{c \Gamma}{kT+c \Gamma}$, and, within the sequential regime, increasing $\Gamma$ increases the degree of violation.
At high bias, the degree of violation becomes less and vanishes in the infinite bias limit. In this limit, a Markovian, essentially classical, description of electron jumps between system and reservoir emerges.

\section{Superconducting SET}
The second SET we consider has superconducting island and leads.  Under applied bias, transport through the island can proceed via a number of different channels involving both coherent tunneling of cooper-pairs (CPs) and incoherent tunneling of quasi-particles\cite{Hadley1998}.
We focus here on the DJQP resonance \cite{Clerk2002, *Clerk2003,Thalakulam2004}, where both coherent and incoherent processes occur at both junctions. 
Recent interest in this resonance concerns the current noise at finite-frequency \cite{Xue2009,Kirton2012}.

As discussed in Ref.~\cite{Kirton2012}, the DJQP resonance can be described by a model in the  basis $\{\ket{n,N}\}$ where $n$ is the number of electrons on the island and $N$ the number of charges in one of the leads 
\footnote{
  There are two differences between our calculation and that of Ref.\cite{Kirton2012}. We write our model in terms of collector charges, and derive our results from MGF \eq{MGFlevitov}, whence the different assignment of counting fields.
}.
The coherent part of the evolution is described by a $\chi$-dependent Hamiltonian $H(\chi)=H_C+H_J(\chi)$, with charging part
\beq
  H_C=\sum_{N,n}
  [E_C (n-n_g)^2 -(N+\textstyle{\frac{n}{2}}) eV] \op{n,N}{n,N}
  ,
\eeq
where $V$ is the bias voltage and $n_g$ is the gate-induced island charge; and the Josephson term
\beq
  H_J(\chi) &=& -\textstyle{\frac{1}{2}}E_J 
  \sum_N
	 \ket{0,N}\bra{2,N} 
   \nonumber\\
   &&
   ~~~~~~~~~
   + e^{-2i\chi}\ket{1,N} \bra{-1,N+2} 
   +\mathrm{H.c}
   ,
\eeq
where $E_J$ is the junction Josephson energy. 
A DJQP resonance occurs at voltages for which $n_g = 1/2$ and $eV =2 E_C$, such that the detunings from both left and right junctions vanish.
Transport through the SSET is then described by the {\em Markovian} master equation 
$
  \dot{\rho}(\chi) = 
  -i \left[ H(\frac{1}{2}\chi)\rho (\chi)- \rho(\chi) H(-\frac{1}{2}\chi) \right]
  +\mathcal{W}_{dec}(\chi)\rho(\chi)
$
where the term
$
  \mathcal{W}_{dec}(\chi)\rho(\chi)
  =\Gamma \sum_{N,\alpha} 
  \left[
    e^{i\chi \delta_{\alpha,R}}
    D^N_{\alpha} \rho(\chi) {D^N_{\alpha}}^\dag 
    \!\!\!
    -\textstyle{\frac{1}{2}}
    \{{D_\alpha^N}^\dag D^N_{\alpha}, \rho(\chi) \}
  \right]
$ with $D^N_{L} = \op{0,N}{-1,N}$ and  $D^N_{R} = \op{1,N+1}{2,N}$, describes dissipative quasiparticle tunneling at a rate $\Gamma$.
The resultant $\chi$-dependent master equation has an $8\times8$ block matrix structure \cite{Clerk2002,*Clerk2003}, which renders the problem tractable without further approximation.

\begin{figure}[tb]
  \includegraphics[width=\columnwidth,clip=true]{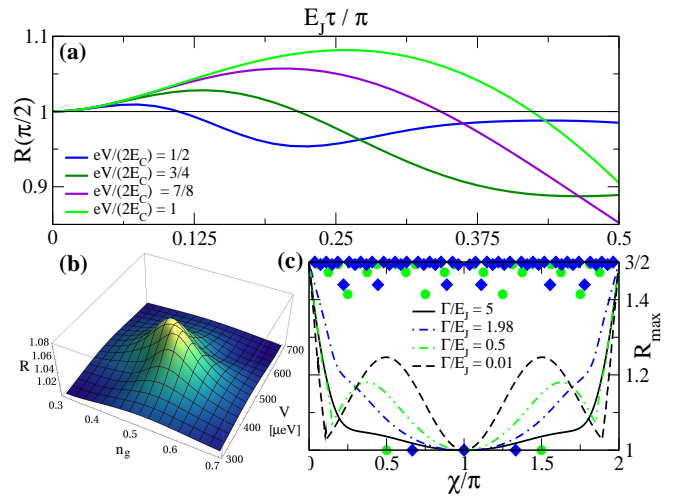}
  \caption{
    (Color online)  Violation of inequality \eq{ineqR} for the superconducting-SET near DJQP resonance.  
    {\bf (a)} $R(\pi/2,\tau)$ as a function of time on resonance. Violations are clearly seen. For small quasi-particle tunneling rate $\Gamma$, the violations repeat, albeit somewhat damped, at longer times.
    {\bf (b)} The maximum $R_\mathrm{max}(\pi/2)$ as a function of gate charge and bias.  DJQP resonance occurs at $n_g=1/2$ and $eV = 474\mu$eV.
    {\bf (c)}  $R_\mathrm{max}$ as a function of $\chi$ on resonance (see text for discussion).
    Unless otherwise stated, the parameters were $\Gamma/E_J=1.98$,  $E_C/E_J = 4.65$, on resonance: $n_g=1/2$ and $eV=2E_C$.
    \label{SSETfig}
   }
\end{figure}

Figure \fig{SSETfig}a shows the real part $R$ as a function of time and \fig{SSETfig}b its maximum value as a function of bias and gate charge for a value $\chi=\pi/2$.  The parameters used were the experimental parameters of Ref.~\cite{Xue2009} and inequality \eq{ineqR} is clearly violated around the DJQP resonance.  

The behaviour of $R_\mathrm{max}(\chi)$ as a function of $\chi$ is interesting (\fig{SSETfig}c).  Unlike for the qubit or the SET, $R_\mathrm{max}(\chi=\pi)$ is zero. This is because there are no coherent processes involving the transfer of single charges here.  
In the limit $\Gamma/E_J \to 0$, CP tunneling dominates quasi-particle tunneling and the systems behaviour is essentially a mixture of CP tunneling in the left junction and CP tunneling in the right. Each process looks like the oscillation of our qubit (with double charge), but only the righthand process contribute to the generating function.  We have then
$R(\chi) =\frac{1}{2}\left[1+L_\mathrm{qb}(2\chi)\right] = 1 + \frac{1}{4}\sin^2(\chi)$, with a maximum value of $\frac{1}{4}$, as observed.
In the incoherent limit, $E_J/\Gamma \to 0$, we have $R(\chi)=1$. At finite $E_J$, however, we observe a peak  with $R_\mathrm{max}(\chi) \to \frac{3}{2}$ for small values of $\chi$ (and for $\chi\to 2\pi$).  In this regime, the MGF looks likes that of qubit with $\chi=\pi$ with oscillation frequency $\Omega_\mathrm{eff}\sim\frac{3}{2}\Gamma E_J^2\chi/(\Gamma^2 + 2 E_J)$.   Thus, $R_\mathrm{max}(\chi) \to \frac{3}{2}$ for small $\chi$.  This does not constitute a violation of \eq{ineqR}, however, because the upper bound $C_\mathrm{R}(\chi)$ in this limit is also $\frac{3}{2}$.

For this model then, the greatest violations are found for $\chi=\frac{1}{2}\pi$.  For the experimental values of Ref.~\cite{Xue2009}, the maximum value is $R_\mathrm{max}\approx 1.08$.  Decreasing the rate $\Gamma$ by one-half, would double the degree of violation, and an order of magnitude would bring the violation close to the theoretical maximum of $\frac{5}{4}$.  These results not only confirm the quantum nature of the DJQP resonance, but also show that no macroscopic-real explanation of this phenomenon is possible.

\section{Discussion}
We have described a set of Leggett-Garg inequalities for the MGF of FCS.  Violation of this inequality constitutes absence of ``macroscopic reality'' of the charge transferred to the reservoir.
We have calculated violations for our quantum systems using the MGF of \eq{MGFlevitov}.  However, similar violations can be found with alternative definitions. For example, the MGF discussed by Shelankov and Rammer \cite{Shelankov2003},
\beq
  \mathcal{G}_\mathrm{SR}(\chi,t) 
  &=& 
  \sum_n
  \ew{
    \hat{P}_n
        e^{-i\frac{\chi}{2} \hat{n}(t_a)}
    e^{i\chi\hat{n}(t_b)}
    e^{-i\frac{\chi}{2} \hat{n}(t_a)} 
    \hat{P}_n
  }
  \label{FCS_MGF_shel}
  ,
\eeq
which includes a sum over projection operators $\hat{P}_n$ onto initial states of definite charge, gives real-part violations that are identical for the qubit
and very similar for the two SET models to those obtained from \eq{MGFlevitov}.
The imaginary parts calculated with this second MGF are, however, always zero, suggesting that the real-part violations are a more robust indicator of quantum behaviour.

One interesting perspective is to see how the uniquely quantum behaviour described here relates to the connection between FCS and entanglement \cite{Song2011}.

\begin{acknowledgments}
I am grateful to N.~Lambert, T.~Brandes and M.~Houzet for discussions and to 
O.~Karlstr\"{o}m and J.~Pederson for help with the SET calculations.  This work was supported by the DFG through SFB 910.
\end{acknowledgments}


\end{document}